\documentclass[aps,graphicx,showpacs,superscriptaddress,amsmath,amssymb]{revtex4}
\usepackage{graphicx}

\begin{document}

\title{Practical scheme for quantum dense coding between three parties using microwave radiation in trapped ions}

\author{Wen-Xing Yang} \email{wenxingyang2@126.com}
\affiliation{Department of Physics, Southeast University, Nanjing 210096, China} \affiliation{State Key Laboratory of Magnetic Resonance and Atomic and
Molecular Physics, Wuhan Institute of Physics and Mathematics, Chinese Academy of Sciences, Wuhan 430071, China}
\author{Zhe-Xuan Gong} \email{gongzhexuan@gmail.com}
\affiliation{Department of Physics, Huazhong University of Science and Technology, Wuhan 430074, China}

\date{\today}

\begin{abstract}
We propose a practical scheme for implementing two-dimension quantum dense coding (QDC) between three parties through manipulating three ions confined in
microtraps addressed by microwaves and assisted by a magnetic field gradient. The ions in our scheme are not required to be strictly cooled to the vibrational
ground state because single-qubit and multi-qubit operations are made via Ising terms, in which the vibrational modes of the ions remain unchanged throughout
the scheme, rendering our scheme robust to the heating of the ions. We also present the detailed steps and parameters for implementing the three-party QDC
experimentally and show that the proposed scheme is within the current techniques of ion-trap experiments.
\end{abstract}


\maketitle

Quantum Dense Coding (QDC) \cite{1,2} is one of the many surprising applications of quantum entanglement in quantum communication. The well-known
two-dimension, two-party QDC can transmit two bits of classical information through a quantum channel by actually sending a single qubit. Briefly, if the
receiver Alice and the sender Bob previously share a Bell-state entanglement pair and then Bob operates locally on his particle one of the four unitary
transformations \{$I$, $\sigma_{x}$, $i\sigma_{y}$, $\sigma_{z}$\} before sending it to Alice, Alice can obtain two bits of classical information via
collective measurement on both particles. Such kind of QDC has been studied extensively for its theoretical properties and security concerns \cite{3,4}, and
realized experimentally by using entangled photons \cite{5,6} and Nuclear Magnetic Resonance(NMR) techniques \cite{7}. Recently, QDC has been a focus again as
it was extended to high dimensions and multi parties, which may give rise to many new interesting applications \cite{8}-\cite{14}.

Multi-party QDC \cite{11,12} has not been carried out experimentally till now. Decoherence might be the main obstacle for the implementation of such kind of
quantum information processing (QIP) that requires scalability and high fidelity. For example, the trapped ion system, which is known to be a qualified
candidate for QIP, is subjected to decoherence that arises mainly from the heating of the trapped ions, and also from imprecise laser and operation instability
of optical frequencies \cite{15}. In this paper, we overcome such obstacles for implementing QDC in trapped ion system by using modified ion traps, in which
the trapped ions interact via spin-spin coupling produced by a magnetic field gradient and Coulomb force \cite{16,17,18}. Single-qubit and multi-qubit
operations are thus made via Ising terms, in which the vibrational modes of the ions remain unchanged throughout the scheme. Therefore we do not require the
ions to be strictly cooled to the vibrational ground state, rendering our scheme robust to the heating of the ions. In contrast to the original ion trap scheme
for QDC \cite{19}, we also make use of the microwave source instead of the Raman-type radiation, which makes practical experiments easier than those using
lasers. Based on such physical model, we present the concrete steps and suitable parameters for implementing each quantum operation in QDC, and our discussion
shows that they are within the current techniques of ion-trap experiments.

\section{Two-dimension, three-party quantum dense coding}
We'd like to introduce multi-party QDC by focusing on the simplest case of two-dimension QDC between three parties. Assume in the following case that Alice is
the only receiver while Bob and Claire are two senders. The three parties previously share a three-qubit GHZ state:
\begin{equation}\label{eq1}
  |\psi\rangle_{ABC}=\frac{1}{\sqrt{2}}(|0\rangle_A |0\rangle_B |0\rangle_C + |1\rangle_A |1\rangle_B |1\rangle_C)
\end{equation}

Then Bob and Claire are to operate locally on their qubits respectively and independently. If their single-qubit unitary operations are chosen appropriately,
$|\psi\rangle_{ABC}$ will change into one of the following 8 three-qubit maximal entangled states:
\begin{equation}\label{eq2}
\left \{
\begin{array}{l}
 \Phi_{000}=(|0\rangle_A |0\rangle_B |0\rangle_C + |1\rangle_A |1\rangle_B |1\rangle_C)/\sqrt{2} \\
 \Phi_{001}=(|0\rangle_A |0\rangle_B |1\rangle_C + |1\rangle_A |1\rangle_B |0\rangle_C)/\sqrt{2} \\
 \Phi_{010}=(|0\rangle_A |1\rangle_B |0\rangle_C + |1\rangle_A |0\rangle_B |1\rangle_C)/\sqrt{2} \\
 \Phi_{011}=(|0\rangle_A |1\rangle_B |1\rangle_C + |1\rangle_A |0\rangle_B |0\rangle_C)/\sqrt{2} \\
 \Phi_{100}=(|0\rangle_A |0\rangle_B |0\rangle_C - |1\rangle_A |1\rangle_B |1\rangle_C)/\sqrt{2} \\
 \Phi_{101}=(|0\rangle_A |0\rangle_B |1\rangle_C - |1\rangle_A |1\rangle_B |0\rangle_C)/\sqrt{2} \\
 \Phi_{110}=(|0\rangle_A |1\rangle_B |0\rangle_C - |1\rangle_A |0\rangle_B |1\rangle_C)/\sqrt{2} \\
 \Phi_{111}=(|0\rangle_A |1\rangle_B |1\rangle_C - |1\rangle_A |0\rangle_B |0\rangle_C)/\sqrt{2}
\end{array} \right.
\end{equation}

Then Bob and Claire should send both their qubits to Alice in quantum channels. Alice then performs a collective measurement to find out which one of the 8
orthogonal states in equation (\ref{eq2}) she gets. Since 8 different results can be encoded as a three-bit classical information, Alice achieves, after
receiving two qubits and conducting a single measurement, in obtaining three bits of classical information which is partially provided by Bob and partially by
Claire. This procedure might have interesting use in quantum communication, \emph{e.g.} if Alice, who has previously let two others help her keep some secret
information separately, then wants to retrieve it in a safe way such that neither Bob nor Claire can know it. Furthermore, it is worth mentioning that when
extended to arbitrary number of parties, Alice can retrieve $N$-bit information always by a SINGLE measurement after receiving $N-1$ qubits from $N-1$ senders.

\section{Our physical model}
\begin{figure}
\includegraphics[width=0.5\textwidth]{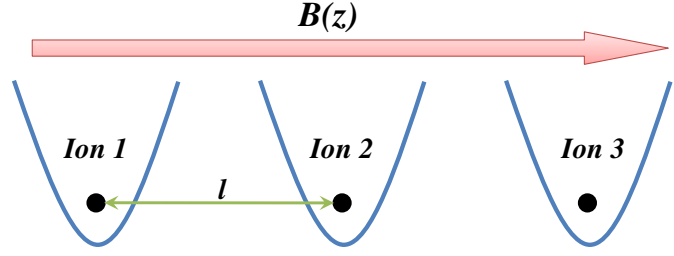}
\caption{Schematic plot for multi-trap model: each ion confined in an individual micro-trap, the magnetic field $B(z)=B_{0}+\frac{\partial B}{\partial z}z$. $l$ is the
separation between neighboring traps.}\label{fig1}
\end{figure}

In our physical model, the atomic hyperfine levels are encoded as qubits. And due to the magnetic field gradient, the ions can be distinguished in frequency
space and individually addressed by microwaves. For pedagogical reasons, our calculations will start from the case of three ions beside each other linearly
confined in three individual traps, based on the multi-trap model in Ref.\cite{18}. The three ions, for example $Yb^+$, are in a magnetic field with gradient
in $z$ direction, as shown in figure \ref{fig1}. If we encode qubits as $| {F = 1,m_F = 1} \rangle \to | 1 \rangle $ and $| {F = 0} \rangle \to | 0 \rangle $,
the Hamiltonian for this system can be written as:
\begin{equation}\label{eq3}
H = \sum\limits_{i = 1}^3 {\textstyle{1 \over 2}\omega _i (z_{0,i} )\sigma _{z,i} } + \sum\limits_{p = 1}^3 {\nu _p a_p^\dag a_p } -\frac{1}{2}J_{13}\sigma _{z,1}\sigma
_{z,3} - \frac{1}{2}J(\sigma _{z,1} \sigma _{z,2} + \sigma _{z,2} \sigma _{z,3} )
\end{equation}

\begin{equation}
\label{eq4} J_{ij} = \sum\limits_{p = 1}^3 {\frac{\hbar }{2m\nu _p^2 }D_{i,p} D_{j,p} \frac{\partial \omega _i (z_{0,i} )}{\partial z}\frac{\partial \omega _j (z_{0,j}
)}{\partial z}}
\end{equation}
where $\omega _i (z_{0,i} )$ is the transition frequency for ion $i$ which depends on the equilibrium position $z_{0,i}$ of the ion in the magnetic field
gradient, $\nu _p $ and $\sigma _{z,i} $ is the $p^{th}$ vibrational frequency and the usual Pauli operator for ion $i$, and $D$ is the expansion coefficient
of the displacement of ion $i$ in terms of the normal mode coordinate. For simplicity, we have adjusted the coupling terms $J_{12} = J_{23} = J$ by simply
setting the frequencies of microtrap for ion 1 and 3 to be equal.

Direct algebra calculation shows that the eigenenergies of the equation (\ref{eq3}) are as in table \ref{tab1}, which can be distinguished from each other.
Besides, we give the carrier transition frequency $\omega_c$ of a certain ion with corresponding states of the other two ions as in table \ref{tab2}. It is
obviously that the carrier transition frequencies for each ion are dependent on the states of the other two ions \cite{20}.

\begin{table}[here]
\caption{The eigenenergies in the space spanned by $| 0 \rangle _1 | 0 \rangle _2 | 0 \rangle _3 $, $| 1 \rangle _1 | 0 \rangle _2 | 0 \rangle _3 $, $| 0 \rangle _1 | 1
\rangle _2 | 0 \rangle _3 $, $| 0 \rangle _1 | 0 \rangle _2 | 1 \rangle _3 $, $| 1 \rangle _1 | 1 \rangle _2 | 0 \rangle _3 $, $| 1 \rangle _1 | 0 \rangle _2 | 1 \rangle
_3 $, $| 0 \rangle _1 | 1 \rangle _2 | 1 \rangle _3 $, $| 1 \rangle _1 | 1 \rangle _2 | 1 \rangle _3 $.} \label{tab1}
\begin{tabular}{c|c}
 \hline Eigenvectors (Basis) & Eigenenergies \\ \hline
 $| 0 \rangle _1 | 0 \rangle _2 | 0 \rangle _3 $ \quad\quad\quad & \quad\quad $- 0.5(\omega _1 + \omega _2 + \omega _3 ) - J- 0.5 J_{13}$ \\
 $| 1 \rangle _1 | 0 \rangle _2 | 0 \rangle _3 $ \quad\quad\quad & \quad\quad $ - 0.5( - \omega _1 + \omega _2 + \omega _3 )+ 0.5 J_{13}$ \\
 $| 0 \rangle _1 | 1\rangle _2 | 0 \rangle _3 $ \quad\quad\quad & \quad\quad $ - 0.5(\omega _1 -\omega_2 + \omega _3 ) + J- 0.5 J_{13}$ \\
 $| 0 \rangle _1 | 0 \rangle _2 | 1 \rangle _3 $ \quad\quad\quad & \quad\quad $ - 0.5(\omega _1 + \omega _2 -\omega _3 )+ 0.5 J_{13}$ \\
 $| 1 \rangle _1 | 1 \rangle _2 | 0 \rangle _3 $ \quad\quad\quad & \quad\quad $0.5(\omega _1 + \omega _2 - \omega _3 )+ 0.5 J_{13}$ \\
 $| 1 \rangle _1 | 0 \rangle _2 | 1 \rangle _3 $ \quad\quad\quad & \quad\quad $0.5(\omega _1 - \omega _2 + \omega _3 ) + J - 0.5 J_{13}$ \\
 $| 0 \rangle _1 | 1 \rangle _2 | 1 \rangle _3 $ \quad\quad\quad & \quad\quad $0.5( - \omega _1 + \omega _2 + \omega _3 )+ 0.5 J_{13}$ \\
 $| 1 \rangle _1 | 1 \rangle _2 | 1 \rangle _3 $ \quad\quad\quad & \quad\quad $0.5(\omega _1 + \omega _2 + \omega _3 ) - J- 0.5 J_{13}$ \\
 \hline
\end{tabular}
\end{table}

\begin{table}[htbp]
\caption{The carrier transition frequency $\omega_{c}$ of a certain ion with corresponding states of the other two ions.} \label{tab2}
\begin{tabular}{c|c|c|c|c|c}
 \hline State for ion 2,3 & $\omega_{c}$ of ion 1 & State for ion 1,3 & $\omega_{c}$ of ion 2 & State for ion 1,2 & $\omega_{c}$ of ion 3\\ \hline
 $| 1 \rangle | 1 \rangle $ & $\omega_{1}-J-J_{13}$ & $| 1 \rangle | 1 \rangle $ & $\omega_{2}-2J$ & $| 1 \rangle | 1 \rangle $ & $\omega_{3}- J- J_{13}$ \\
 $| 1 \rangle | 0 \rangle $ & $\omega_{1}-J+ J_{13}$ & $| 1 \rangle | 0 \rangle $ & $\omega_{2}$ & $| 1 \rangle | 0 \rangle $ & $\omega_{3}+J-J_{13}$ \\
 $| 0 \rangle | 1 \rangle $ & $\omega_{1}+J- J_{13}$ & $| 0 \rangle | 1 \rangle $ & $\omega_{2}$ & $| 0 \rangle | 1 \rangle $ & $\omega_{3}-J+J_{13}$ \\
 $| 0 \rangle | 0 \rangle $ & $\omega_{1}+J+J_{13}$ & $| 0 \rangle | 0 \rangle $ & $\omega_{2}+2J$ & $| 0 \rangle | 0 \rangle $ & $\omega_{3}+J+J_{13}$ \\
 \hline
\end{tabular}
\end{table}

In our scheme, to resonantly excite one ion irrespective of the states of the other two ions, we should have a microwave with Rabi frequency much larger than the maximum
difference between the carrier transition frequencies regarding this ion \cite{16}. Assuming the trap frequencies for microtraps of ion 1, ion 2 and ion 3 are 0.5MHz,
5MHz, and 0.5MHz respectively, the separation of neighboring traps $l = 5\mu m$ and the magnetic field gradient $\partial B /\partial z = 200T / m$, we can estimate $J
\approx 10KHz$, much larger than the coupling strength between ion 1 and 3, about 2.0KHz, and the three vibrational mode frequencies are approximately 0.39MHz, 0.64MHz,
and 0.72MHz, and the neighboring qubit resonance frequency separation as $g\mu _B \frac{\partial B}{\partial z}l \approx 163MHz$.

Due to the microwave added, we know the interaction Hamiltonian differs from that in the normal linear trap only by replacing the Lamb-Dicke parameter $\eta
_{i,p} $ with $\eta '_{i,p} = \sqrt {\eta _{i,p}^2 + \varepsilon _{i,p}^2 } $, where the subscript $i$ denotes ion $i$, $p$ represents the $p^{th}$ collective
motional mode, $\eta _{i,p} $ is about $0.7\times10^{ - 6}$. In the Paschen-Bach limit, the frequency gradients are independent of $z$, therefore $\varepsilon
_{i,p} = D_{i,p} (\sqrt {\hbar / 2m\nu _p } \partial B /\partial z) / \nu _p $. With the parameters given in the above example, the maximum is about 0.013, so
all the $\eta '_{i,p} $ are much smaller than 1. Direct calculation shows that under the restriction $\varepsilon_{max}< 0.05$ ($\varepsilon_{max}$ denotes the
maximum of all $\varepsilon$), the smaller the neighboring trap distance $l$, the bigger the $J$ obtainable. Besides, it can be easily shown that the evolution
operator for addressing ion $i$ by a microwave in the interaction picture is,
\begin{equation}
\label{eq5} U_I^i (\theta ,\phi ) = ( {{\begin{array}{*{20}c}
 {\cos \frac{\theta }{2}} \hfill & { - ie^{ - i\phi }\sin \frac{\theta }{2}}
\hfill \\
 { - ie^{ + i\phi }\sin \frac{\theta }{2}} \hfill & {\cos \frac{\theta }{2}}
\hfill \\
\end{array} }} )
\end{equation}
where $\theta = \Omega t$ with the Rabi frequency $\Omega $ being in the order of MHz, and phase $\phi$ relating to the position of ion in the microwave.
Equation (\ref{eq5}) achieves the single qubit-rotation operation that is needed in the implementation of QDC protocol.

To further attain the two-qubit controlled operation, we borrow the technique of realizing Controlled-NOT gate from NMR quantum computing \cite{21}, where the CNOT
operation is achieved through a sequence of pulses. In product operator representation, that is:
\begin{equation}
\label{eq6}
\begin{array}{l}
U_{CNOT} (i,j) = e^{ - i\pi / 4}e^{ - i\pi / 4\sigma _{y,j} }e^{i\pi / 4\sigma _{z,i} }e^{i\pi / 4\sigma _{z,j} }\\\\\quad\quad\quad\quad\quad\quad \times e^{ - i\pi /
4\sigma _{z,i} \sigma _{z,j} }e^{i\pi / 4\sigma _{y,j} }
\end {array}
\end{equation}
where ion $i$ and $j$ act as the controlling qubit and target qubit respectively. $e^{ - i\pi / 4\sigma _{z,i} \sigma _{z,j} }$ can be realized through the
coupling term $\frac{1}{2}J_{i,j} \sigma _{z,i} \sigma _{z,j} $ in equation (\ref{eq3}). The undesired evolution brought by the other terms in equation
(\ref{eq3}) can be eliminated by the refocusing techniques \cite{2}. All the other terms can be implemented by microwave pulses.

In table \ref{tab3} we give a concrete example of $U_{CNOT}(1,2)$. Note that in the context of NMR, Rabi frequency is much larger than other characteristic
frequencies. In this case, $\omega _i \sim 13GHz$ is the biggest frequency, much larger than other characteristic frequencies. Therefore, to avoid undesired
evolution due to $\omega _i (z_{0,i} )$, we have to carefully control the pulse length $t$ to satisfy $\omega _i (z_{0,i} )t = 2n\pi $, ($n = 1,2,3,\ldots )$,
which can be realized by properly adjusting $B_0 $, $\partial B / \partial z$ and $\Omega $. Given the parameters in table \ref{tab3} for $l=5\mu m$, the total
time for completing the CNOT gate is about 2.82ms. Similar parameters can be chosen to realize $U_{CNOT}(2,3)$ too.

\begin{table}[here]
\caption{The pulse sequences and estimated time for implementing each term in $U_{CNOT}(1,2)$, with $U_{0}$ being the unitary evolution operator of the
Hamiltonian in equation (\ref{eq3}) without the second term, and $t_{0}$ being the implementation time for $U_{0}$. Assume the implementation time for any
single qubit rotation $U_{I}^{i}(\theta,\phi)$ to be $5\mu s$. $J$ is at the order of kHz, which could be achieved by properly adjusting the Rabi frequency of
the microwave.} \label{tab3}
\begin{tabular}{c|c|c}
 \hline Term & Pulse sequences & Estimated time \\ \hline
 $e^{i\pi/4\sigma_{y,1}}$ & $U_{I}^{1}(\frac{\pi}{2},\frac{\pi}{2})$ & $5\mu s $\\ \hline
 $e^{-i\pi/4\sigma_{z,1}\sigma_{z,2}}$ &\begin{tabular}{c}
  $U_{I}^{1}(\pi,0)U_{I}^{2}(\pi,0)U_{0}(\frac{t_{0}}{4})$ \\
  $U_{I}^{3}(\pi,0)U_{0}(\frac{t_{0}}{4})U_{I}^{1}(\pi,0)$ \\
  $U_{I}^{2}(\pi,0)U_{0}(\frac{t_{0}}{4})U_{I}^{3}(\pi,0)$ \\
  $U_{0}(\frac{t_{0}}{4})(t_{0}=\frac{7\pi}{2J})$ \\
  \end{tabular} & $2.78ms$ \\ \hline
 $e^{i\pi/4\sigma_{z,1}}$ & $U_{I}^{1}(\frac{\pi}{2},\frac{\pi}{2})U_{I}^{1}(\frac{\pi}{2},0)U_{I}^{1}(\frac{7\pi}{2},\frac{\pi}{2})$ & $15\mu s$ \\ \hline
 $e^{i\pi/4\sigma_{z,2}}$ & $U_{I}^{2}(\frac{\pi}{2},\frac{\pi}{2})U_{I}^{2}(\frac{\pi}{2},0)U_{I}^{2}(\frac{7\pi}{2},\frac{\pi}{2})$ & $15\mu s$ \\ \hline
 $e^{-i\pi/4\sigma_{y,1}}$ & $U_{I}^{1}(\frac{7\pi}{2},\frac{\pi}{2})$ & $5\mu s $ \\ \hline
\end{tabular}
\end{table}

\section{The implementation}
Suppose the qubits hold by Alice, Bob and Claire are represented by ion 1, 2, 3 respectively, which are initially in the state $|0\rangle_1 |0\rangle_2
|0\rangle_3$. Then we first need to prepare the three-qubit GHZ state in equation (\ref{eq1}), which can be attained by first applying a Hadamard operation on
ion 1, followed by a CNOT operation on ion 1, 2, and finally a CNOT operation on ion 2, 3:
\begin{eqnarray}\label{eq7}
|0\rangle_1 |0\rangle_2 |0\rangle_3 &\xrightarrow{H(1)}& \frac{1}{\sqrt{2}}(|0\rangle_1 + |1\rangle_1) |0\rangle_2 |0\rangle_3 \\
&\xrightarrow{U_{CNOT}(1,2)}& \frac{1}{\sqrt{2}}(|0\rangle_1 |0\rangle_2 + |1\rangle_1 |1\rangle_2)|0\rangle_3 \\
&\xrightarrow{U_{CNOT}(2,3)}& \frac{1}{\sqrt{2}}(|0\rangle_1 |0\rangle_2 |0\rangle_3 + |1\rangle_1 |1\rangle_2 |1\rangle_3)
\end{eqnarray}

Next comes the problem of which set of specific single-qubit unitary operations in equation (\ref{eq5}) Bob and Claire should apply to their qubits. We
generally prefer to use the set of Pauli operators $\{I,\sigma_x,i\sigma_y,\sigma_z\}$, but if Bob and Claire can both choose each of the four Pauli operators,
there will be $4\times4=16$ different results, not in accordance with the 8 orthogonal resultant states in equation (\ref{eq1}). In order to prevent confusion
in the information Alice finally obtained, here we prescribe that Bob can perform each of $\{I,\sigma_x,i\sigma_y,\sigma_z\}$ but Claire can only perform $I$
and $\sigma_x$, such as in Ref. \cite{11}. Note that there are also many other practicable choices, especially when extended to arbitrary number of parties
\cite{12}. For three-party case, the choice can only be asymmetric, since Bob actually provides two bits of classical information while Claire provides only
one bit. We show in table \ref{tab4} how to use our physical model to implement those single-qubit unitary operations so as to obtain each of the states in
equation (\ref{eq1}).

\begin{table}
\caption{The state obtained after possible operations through pulse sequences by Bob and Claire. We adopt the same parameters as in table \ref{tab1}.}
\label{tab4}
\begin{tabular}{c|c|c|c}
 \hline State obtained & Operations by Bob and Claire & Pulse Sequences & Estimated time \\ \hline
 $\Phi_{000}$ & $I,I$ & No Pulses & 0 \\
 $\Phi_{001}$ & $I,\sigma_{x}$ & $U_{I}^{3}(\pi,0)$ & $5\mu s$ \\
 $\Phi_{010}$ & $\sigma_{x},I$ & $U_{I}^{2}(\pi,0)$ & $5\mu s$ \\
 $\Phi_{011}$ & $\sigma_{x},\sigma_{x}$ & $U_{I}^{3}(\pi,0),U_{I}^{2}(\pi,0)$ & $10\mu s$ \\
 $\Phi_{100}$ & $\sigma_{z},I$ & $U_{I}^{2}(\pi,\frac{\pi}{2})U_{I}^{2}(\pi,0)$ & $10\mu s$ \\
 $\Phi_{101}$ & $\sigma_{z},\sigma_{x}$ & $U_{I}^{2}(\pi,\frac{\pi}{2})U_{I}^{2}(\pi,0),U_{I}^{3}(\pi,0)$ & $15\mu s$ \\
 $\Phi_{110}$ & $i\sigma_{y},I$ & $U_{I}^{2}(\pi,\frac{\pi}{2})$ & $5\mu s$ \\
 $\Phi_{111}$ & $i\sigma_{y},\sigma_{x}$ & $U_{I}^{2}(\pi,\frac{\pi}{2}),U_{I}^{3}(\pi,0)$ & $10\mu s$ \\
 \hline
\end{tabular}
\end{table}

The final step is to collectively measure all of the three ions to find out the encoded information. The 8 maximal entangled basis in equation (\ref{eq1}) are
not easy to distinguish experimentally, but we can apply the following operations to change them into computational basis:
\begin{equation}
\label{eq6} |\Phi_{abc}\rangle \xrightarrow{U_{CNOT}(2,3),U_{CNOT}(1,2),H(1)} |abc\rangle
\end{equation}
where $a,b,c \in \{0,1\}$ denote the encoded three-bit classical information.

\begin{figure}[here]
\begin{center}
\includegraphics[width=0.5\textwidth]{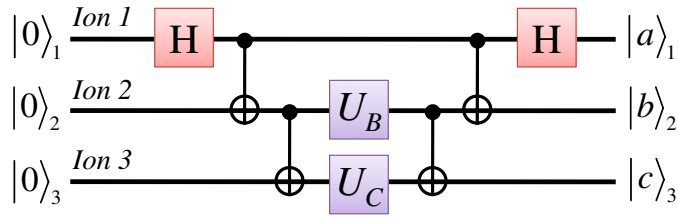}
\caption{Circuit model of the our scheme for implementing the two-dimension QDC between three parties. $H$ denotes the Hadamard gate, $U_{B}$ and $U_{C}$ stand
for the single-qubit unitary operations Bob and Claire perform on the qubits at their hands.}\label{fig2}
\end{center}
\end{figure}

Figure \ref{fig2} clearly depicts the whole procedure of our scheme in circuit model. Alice conducts a single measurement at the output of the circuit and
obtains the encoded information. Till now, we have already shown how to implement all the operations needed in the our whole procedure.

\section{Discussions}
It is necessary to give some brief discussions on the experimental feasibility of our scheme. Compared with the Cirac-Zoller gate operation \cite{22} and the
geometric phase gate operation \cite{19,23}, our scheme does not require moving or hiding any ions, which is important in view of decoherence \cite{24}. Based
on the table \ref{tab3}, we estimate the two-qubit CNOT gate operation time to be 3 ms. As an example, we supposed 10\% of the modes are really excited during
our implementation of the gate, then our scheme would work as long as the heating times of the vibrational states are longer than 0.3 ms. which meets the
requirement of current technology. Besides, the microtrap in our consideration is of the size of 5$\mu m$, and experimentally, three ions in a linear trap with
spacing of the order of $\mu m$ has already been achieved \cite{22}. In addition, microwave with a certain bandwidth and a rapid change of phase is a perfect
technique and magnetic field gradient up to 8000T/m is within the reach of current experiments \cite{25}. In-trap magnetic field gradient of 200 T/m over
several micrometers, which is used in our scheme, can be easily achieved through methods similar to those presented in Ref. \cite{26,27}.

Finally, as previously shown by Wunderlich \cite{17}, the magnetic field gradient allows the qubits to be individually addressed in the microwave via Zeeman
splitting of the hyperfine structure, and the qubit readout process can be achieved through trapped ion fluorescence shelving techniques. We know that the
fluorescence detecting measurement takes 250$\mu s$ \cite{22} base on the parameters in table \ref{tab3}, \ref{tab4}, thus we can estimate that the whole
quantum dense coding process takes less than 6ms. The implementation time can be further shortened by increasing the spin-spin coupling strength $J$, which can
be done by reducing the inter-trap spacing $l$ or enlarging the magnetic field gradient.

Our scheme is implemented in linear ion traps, whereas it will also be practicable in multi-trap devices considering currently available techniques \cite{22}.
Thus our scheme can in principle be extended to QDC between arbitrary number of parties, as theoretically proposed in Ref. \cite{11,12}, where the maximal
entangled state can still be prepared through Hadamard and CNOT gates, and the Pauli operations each sender performs can be carried out using the same
evolution operator as in equation (\ref{eq5}). It is our hope that our scheme can guide future experiment for these QDC procedures, as well as stimulating
further discourse on related QIP techniques in real physical systems.

\begin{acknowledgements}
The authors would thank Prof. Ying Wu for many enlightening discussions and helpful suggestions. This work was partially supported by National Fundamental
Research Program of China 2005CB724508 and by National Natural Science Foundation of China under Grant Nos. 60478029, 90503010, 10634060 and 10575040.
\end{acknowledgements}

\end{document}